\documentclass[prl,twocolumn,showpacs,floatfix,preprintnumbers,amsmath,amssymb,superscriptaddress]{revtex4}

\usepackage{graphicx}
\usepackage{dcolumn}
\usepackage{bm}
\usepackage{color}
\usepackage{color}
\usepackage[latin1]{inputenc}
\usepackage[urlcolor=blue]{hyperref}
\hypersetup{backref, colorlinks=true, linkcolor=blue, citecolor=blue}

\begin{document}

\title{{Signature of multigap nodeless superconductivity in fluorine-doped NdFeAsO}}

\author{A. Adamski}
\affiliation{Institute of Physics, Goethe University Frankfurt, 60438 Frankfurt/M, Germany}

\author{C. Krellner}
\affiliation{Institute of Physics, Goethe University Frankfurt, 60438 Frankfurt/M, Germany}

\author{M. Abdel-Hafiez}
\affiliation{Institute of Physics, Goethe University Frankfurt, 60438 Frankfurt/M, Germany}
\affiliation{Center for High Pressure Science and Technology Advanced Research, Beijing, 100094, China}

\date{\today}

\begin{abstract}
We investigate the temperature dependence of the lower critical field $H_{c1}(T)$, the field at which vortices penetrate into the sample, of a high-quality fluorine-doped NdFeAsO single crystal under static magnetic fields $H$ parallel to the $c$-axis. The temperature dependence of the first vortex penetration field has been experimentally obtained and pronounced changes of the $H_{c1}$(T) curvature are observed, which is attributed to the multiband superconductivity. Using a two-band model with $s$-wave-like gaps, the temperature-dependence of the lower critical field $H_{c1}(T)$ can be well described. These observations clearly show that the superconducting energy gap in fluorine-doped NdFeAsO is nodeless. The values of the penetration depth at $T$ = 0\,K have been determined and confirm that the pnictide superconductors obey an Uemura-style relationship between $T_{c}$ and $\lambda_{ab}(0)^{-2}$.
\end{abstract}

\pacs{74.20.Rp, 74.25Ha, 74.25.Dw, 74.25.Jb, 74.70.Dd}

\maketitle



Superconductivity in the iron-pnictide family has been studied intensively due to the comparably large transition temperatures $T_\mathrm{c}$ of up to 55~K, the unconventional superconducting (SC) properties and interplay with various electronic ground states, such as nematic phase and magnetism~\cite{ZA,ZA1,ZA3,ZA4,Johnston2010,Paglione2010}. One of the crucial issues in understanding the SC mechanism in pnictides is the pairing symmetry of the SC gap~\cite{gap,gap1}. Although there is a general consensus that spin fluctuations play an important role in the formation of Cooper pairs in pnictides, many aspects such as the role of magnetism, the nature of chemical tuning, and the resultant pairing symmetry remain unsettled~\cite{SC,Hir}. Recently, Chubukov and Hirschfeld have shown that there is no general consensus on the nature of pairing in iron-based superconductors leaving the perspectives ranging from $s^{++}$ wave, to $s^{\pm}$, and $d$-wave~\cite{Hir2}. It is quite different from that of cuprates in which almost all have a nodal pairing state~\cite{Cup}. Different scenarios have been proposed to explain the mechanism of the superconductivity in pnictides which pointed to the existence of two-gap, isotropic and anisotropic $s$-wave, $d$-wave and even $p$-wave mechanism~\cite{two,two1,two2,two3,Nd1,Nd2,P,P1,P2,P3,P4}. Such scattered pairing symmetries and various interpretations occur partly due to a sensitive dependence on measurement probes and material quality and stoichiometry.




In view of the existing divergence of conclusions about the gap symmetry, there is a clear need to obtain a set of data by comprehensive study. Lower critical field ($H_{c1}$), or equivalently, magnetic penetration depth ($\lambda$) is  an excellent tool to address this question. The $H_{c1}$, i.e., the thermodynamic field at which the presence of vortices into the sample becomes energetically favorable, is a very useful parameter, providing key information regarding bulk thermodynamic properties and carrying information about the underlying pairing mechanism. Indeed, the gap properties of different families of pnictides have been investigated by tracking the $H_{c1}$ and the magnetic penetration depth~\cite{m2,m3,m4,CR,REN,Martin,RG}. The gap properties of these different families display single to double gaps and even the presence of nodes. Up to now, there have been few investigations on the pairing symmetry of NdFeAsO$_{1-x}$F$_{x}$ system. For instance, penetration depth and point-contact
andreev-reflection spectroscopy show the absence of nodes in the SC gap~\cite{Nd1,Nd2}. The presence of both isotropic and anisotropic $s$-wave symmetry of the order parameter in NdFeAsO$_{0.9}$F$_{0.1}$ has been proposed based on angle resolved photoemission spectroscopy studies~\cite{Nd3}. In contrast to that, $H_{c1}$ experiment on the polycrystalline sample, NdFeAsO$_{0.82}$F$_{0.18}$,
indicates no $s$-wave superconductivity but, rather, nodal gap structure~\cite{Nd4}. However, the non-s-wave-like behavior may come from the granularity of the multigrain oxypnictites in the polycrystalline sample. Clearly, to date, further measurements to elucidate the origin of the SC pairing mechanism are necessary.


Given this background, in this Rapid Communication, we report the results of magnetic and transport studies to resolve the currently debated issue on SC pairing symmetry by using high-quality single crystals in fluorine-doped NdFeAsO. Based on the comprehensive low-$T$ measurements of the $H_{c1} (T)$, a kinky structure in $H_{c1} (T)$ is obtained, which gives strong evidence for two energy gaps, which implies that several sheets of the Fermi surface contribute to the formation of Cooper pairs. Both energy gaps fall with temperature in the way different from the single-band BCS-like behavior.


\begin{figure*}[tbp]
\includegraphics[width=37pc,clip]{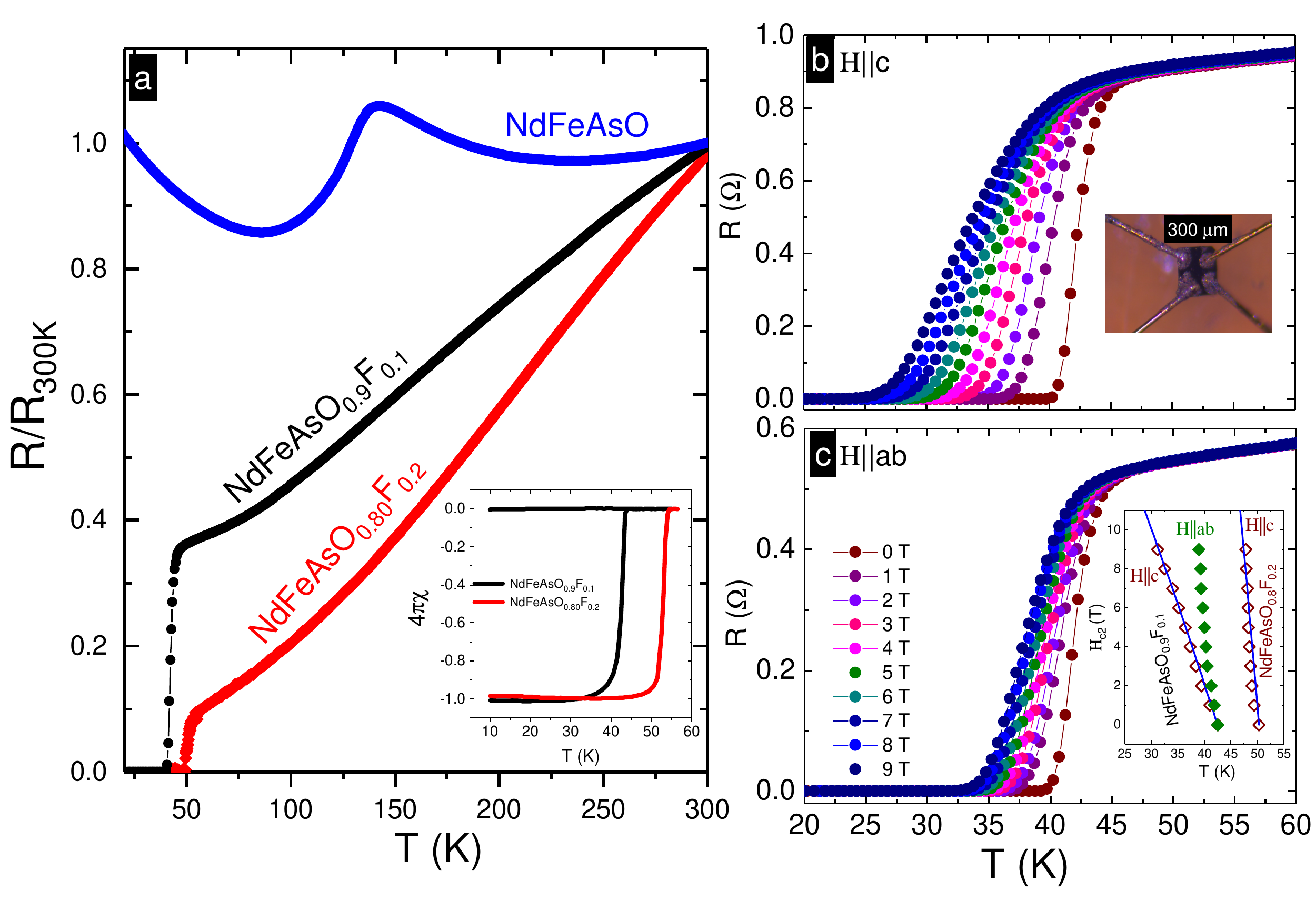}
\caption{\label{fig:wide}  (a) illustrates the $T$-dependence of the in-plane resistance measurements upon heating of NdFeAsO$_{1-x}$F$_{x}$ single crystals or $x$ = 0, 0.1 and 0.2. The insets presents the magnetic susceptibility of for $x$ = 0.1 and 0.2 samples measured in an external field of 1\,Oe applied along $c$-axis. (b) and (c) summarize the $T$-dependent resistance measured in various magnetic fields $x$= 0.1. The inset of (c) demonstrating the four contacts at the surface. The inset of (c) illustrate the phase diagram of $H_{\mathrm{c2}}$ vs.~temperature of $x$= 0.1 and 0.2 for the field applied parallel to $c$. The solid lines are fits to the WHH model for $\lambda = 0$ with $\alpha = 0$.}
\end{figure*}

Single crystals of NdFeAsO and fluorine-doped NdFeAsO were grown out of NaCl/KCl flux at ambient pressure as a new method. All preparation steps like weighing, mixing, and storage were carried out in an Ar-filled glove-box, the O$_{2}$ and H$_{2}$O level is less than 1\,ppm. Starting materials were pristine Neodymium and arsenic as well as iron(III)-oxid powder and iron(II)-fluoride powder. After mixing together the educts were transferred into a glassy garbon crucible. As flux material an eutectic mixture of NaCl and KCl was used, with a molar material to flux ratio of 1:7. The crucibles were sealed into silica ampoules. Because of the vapour pressure of arsenic a slow heating rate of 30\,K/h was used till the maximum temperature of 1473\,K was reached. After a dwell time of 2\,h the ampoules were cooled down to 1273\,K with a rate of 5\,K/h, followed  by a cooling rate of 2\,K/h till 1073\,K. Flux was removed by washing the sample with distilled water. The crystals were collected by filtration. Batches with platlet-like single crystals up to (200$\times$200)$\mu$m$^{2}$ and masses up to 0.2\,mg were carefully examined by Superprobe electron probe microanalyzer and x-ray powder diffraction.

\begin{figure}[b]
\includegraphics[width=20pc,clip]{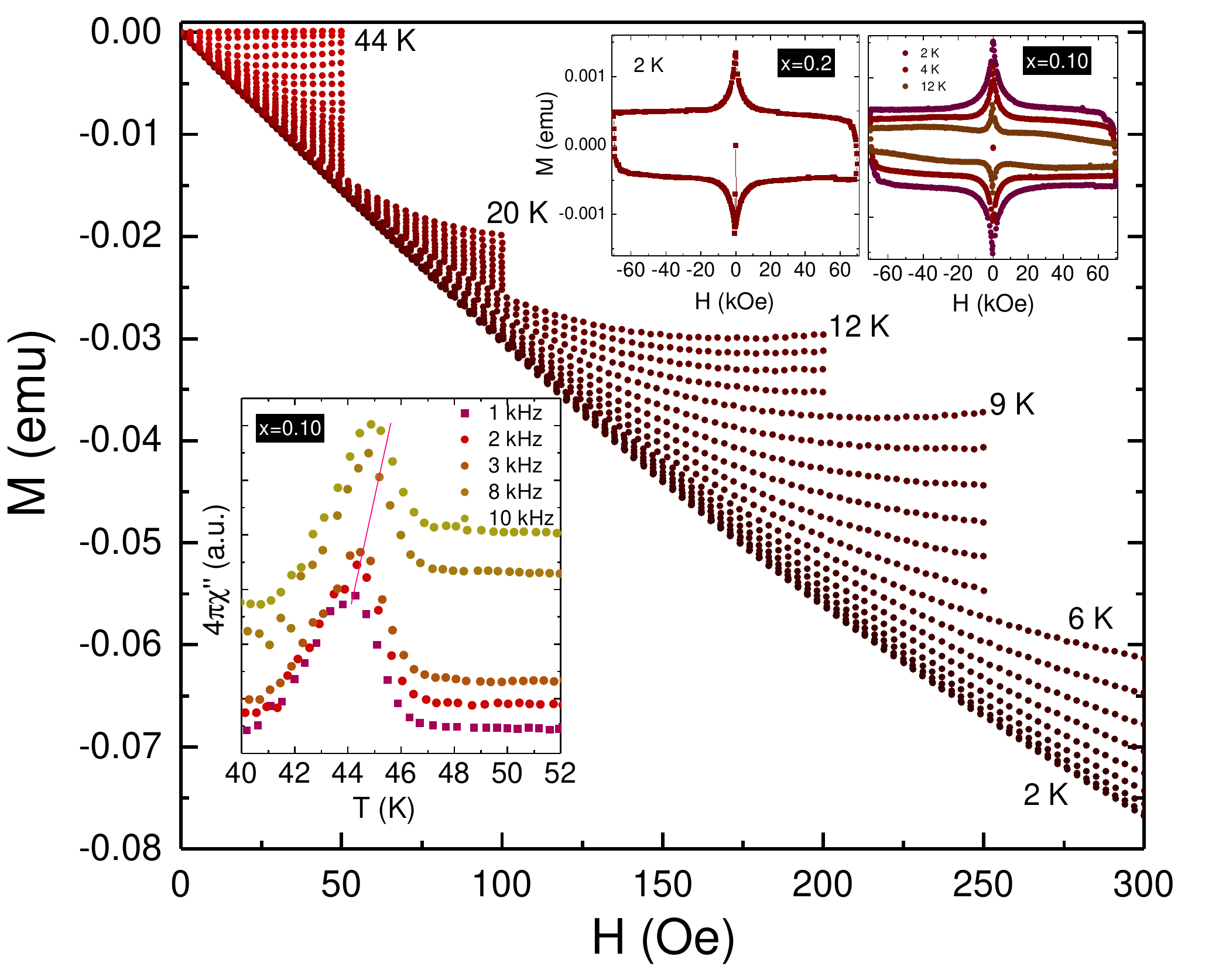}
\caption{The main panel presents the SC initial part of the magnetization curves measured of $x$ = 0.1 single crystals at various temperatures for $H\parallel c$. The upper insets depict the magnetic field dependence of the isothermal magnetization $M$ vs. $H$ loops measured at different temperatures ranging from 2 to 12\,K up to 70\,kOe with the field parallel to  $c$ axis for $x$ = 0.1 and 2\,K for $x$ = 0.2. The lower inset illustrates the imaginary part of ac susceptibility at various frequencies.}
\end{figure}

Figure 1(a) illustrates the $T$-dependence of the in-plane resistance measurements upon heating of NdFeAsO single crystals. Obviously, the anomaly at $T_{S} \approx$ 142\,K in resistance measurements correlates with the phase transition from tetragonal $P$4/$nmm$ to orthorhombic $C_{mma}$. The second anomaly found at $T_{N} \approx$ 130\,K corresponds to the onset of the well known spin density wave (SDW) stripe like Fe order~\cite{1a}. The transitions in NdFeAsO system are consistent with the values obtained from previous reports~\cite{1b} although the crystals were grown with a different method, as described above. F-doped samples show a typical cool-down resistance curve for NdFeAsO$_{1-x}$F$_{x}$ ($x$= 0.1 and 0.2) single crystals in the $T$-range 2-300\,K for current flowing within the planes in absence of an externally applied field. In the F-doped sample, the SDW/structural anomaly is completely suppressed. The normal state exhibits simple metallic behavior upon cooling down from room-$T$, followed by a sharp SC transition at $T_{c} \approx$ 45.4\,K and 52.8\,K (90$\%$ of the normal state resistivity) with $\Delta T_{c}$ = 0.15\,K and 0.3\,K  for $x$ = 0.1 and 0.2 respectively, which is in agreement with the magnetization data, see below. The residual resistivity ratio (RRR) is found to be 3.2 and 10.3 for $x$ = 0.1 and 0.2 respectively. The observations of a large RRR and a narrow SC transition again indicate a high quality of the samples investigated here. Figs.\,1(b) and 1(c) summarize the $T$-dependent resistance measured in various magnetic fields for the NdFeAsO$_{1-x}$F$_{x}$ ($x$= 0.1) sample. Under a magnetic field of 9\,T, the SC transition is broadened and shifted significantly to a lower temperature (see also supplementary materials). In order to illustrate the fact that our investigated samples indicate a very high upper critical field $H_\textup{c2}$, we show the magnetic phase diagram in the lower inset of Fig.\,1(c). $T_\mathrm{c}$ of $x$= 0.1 has been estimated from the resistance experiment, while $T_\mathrm{c}$ of $x$= 0.2 obtained from the ac measurements. The Werthamer-Helfand-Hohenberg (WHH) theory predicts the behavior of $H_\mathrm{c2}(T_\mathrm{c})$ taking into account paramagnetic and orbital pair-breaking~\cite{NR}. The orbital limiting field $H_{\mathrm{c2}}^{\mathrm{orb}}$ at zero temperature is determined by the slope at $T_\mathrm{c}$ as $H_{\mathrm{c2}}^{\mathrm{orb}} = 0.69 \,T_{\mathrm{c}}\, (\partial H_{\mathrm{c2}}/\partial T) |_{T_{\mathrm{c}}}$. Using fit to the data in the whole measurement range for negligible spin-paramagnetic effects ($\alpha = 0$) and spin-orbit scattering ($\lambda = 0$) yields $\mu_0 H_{\mathrm{c2}}^{\mathrm{orb}}(B \parallel c) = $65 and 95~T for $x$ = 0.1 and 0.2 respectively. In reality, paramagnetic and spin-orbit effects are expected to play a role. However, a fit to the data including $\alpha$ and $\lambda$ as free parameters is not reasonable in our case due to the limited field range of investigation. From the $H_{c2}$ values we can estimate the Ginzburg-Landau coherence lengths to be $\xi _{ab} = 4.2$~nm and $\xi _{ab} = 3.9$~nm for $x$ = 0.1 and 0.2 respectively. The mean free path (determined from the resistivity at 45\,K and 50\,K for $x$= 0.1 and 0.2) is estimated to be $\approx$\,12.5\,nm and 15\,nm, respectively. The values of the mean free path are much larger than the obtained coherence length for $x$= 0.1 and 0.2, placing the investigated systems in the clean limit. The derived superconducting parameters are summarized in Table I for both investigated samples for $H \parallel c$. Interestingly, for $x$ = 0.1, the $H_{c2}$ is estimated for $H\parallel ab \approx$ 290\,T, which this value clearly exceeds the Pauli limit. Further information about the anisotropy of $x$ = 0.1 can be obtained, which is $\Gamma=H_{c2}^{(ab)}/H_{c2}^{(c)} \sim$ 4.5. This value is in good agreement with $\Gamma$-values of e.g. NdFeAsO$_{0.82}$F$_{0.18}$~\cite{Jia2008} and SmFe$_{1-x}$Co$_{x}$AsO~\cite{sm}.

\begin{table*}
\caption{\label{tab:table 1} The derived SC parameters for NdFeAsO$_{1-x}$F$_{x}$ along $H \parallel c$}.
\begin{ruledtabular}
\begin{tabular}{ccccccccc}
NdFeAsO$_{1-x}$F$_{x}$ &$T_c$ (K) &$-\frac{d\mu_{0}H_{c2}}{dT}|_{Tc}$ (T/K) &$\mu_{0}H_{c2}$ (T) &$H_{c1}$ (Oe)(0K)& $\ell$ (nm) & $\lambda$ (nm) & $\xi$ (nm)&  $\Delta_{1}, \Delta_{2}$ (meV) \\

\hline
$x$=0.1 &45.4 &2.23  &65 &250 &12.5 &218(10) & 4.2 & 2.2, 11.5\\

\hline
$x$=0.2 &52.8 &3.12  &95 &302 &15 &208(10) & 3.9& 2.8, 12.9 \\

\end{tabular}
\end{ruledtabular}
\end{table*}


The inset of Fig.\,1(a), presents the $T$-dependence of the magnetic susceptibility of both F-doped single crystals measured by following zero-field cooled (ZFC) and field-cooled (FC) procedures in an external field of 1\,Oe applied along $c$-axis. The magnetic susceptibility exhibits a SC transition with an onset transition temperature $T_\mathrm{c}^{\chi}$ of 44.5\,K and 52.5\,K for $x$= 0.1 and 0.2, respectively. The clear irreversibility between FC and ZFC measurements is consequence of a strong vortex trapping mechanism, either by surface barriers or bulk pinning. The shape of each magnetization loop (MHL) of both samples is almost symmetric about the horizontal axis. This indicates that the hysteresis in the crystal arises mainly from bulk flux pinning rather than from the surface barrier. It is worth noting that the SC MHL can still be measured at temperatures very close to $T_{c}$ with only a weak magnetic background. This indicates that the sample contains negligible magnetic impurities. The lower inset in Fig.\,2 presents the imaginary part of ac at various frequencies for $x$ = 0.1. One can clearly see that the peak maxima shifts to a higher temperatures upon increasing the frequency which is apparently due to the motion of vortices.


Determining the $H_{c1}$ from magnetization measurements has never been an easy task, particularly since $H_{c1}$ is an equilibrium thermodynamic field, whereas the magnetization curve is highly irreversible as a consequence of metastable states far from equilibrium~\cite{Moshchalkov,Angst}. Once the values of $H_{c1}$ have been experimentally determined (see supplementary part), we need to correct them accounting for the demagnetization effects. Indeed, the deflection of field lines around the sample leads to a more pronounced Meissner slope given by $M/H_{a} = -1/(1-N)$, where $N$ is the demagnetization factor. Taking into account these effects, the absolute value of $H _{c1}$ can be estimated by using the relation proposed by Brandt:~\cite{Brandt}


\begin{equation}
\label{eq2} q_{disk} = \frac{4}{3\pi}+\frac{2}{3\pi}{\tanh[1.27\frac{b}{a}\ln(1+\frac{a}{b})]},
\end{equation}
where $q \equiv (|M/H_{a}|-1)(b/a)$, and $a$ is the average of the dimensions perpendicular to the field of our investigated sample. For our sample we find $N$ $\approx 0.97$, which is assumed for the actual rectangular shape of our investigated sample.

\begin{figure}[b]
\includegraphics[width=21pc,clip]{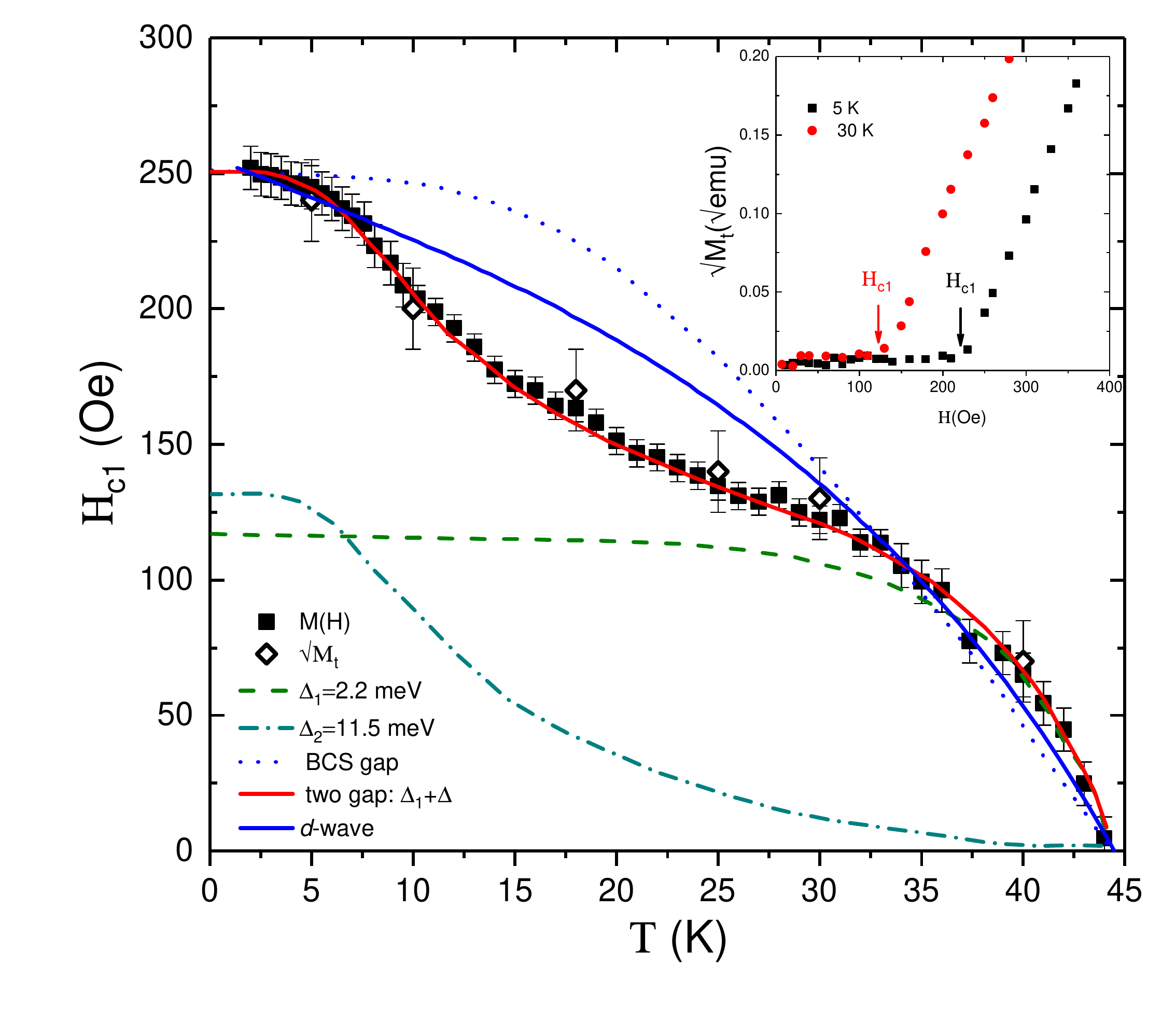}
\caption{Phase diagram of $H_{\mathrm{c1}}$ vs. $T$ for the field applied parallel to $c$ axis. $H_\mathrm{c1}$ has been estimated by two different methods - from the extrapolation of $\sqrt{M_{t}}\rightarrow 0$ (open symbols, see upper inset) and from the regression factor (closed symbols). The bars show the uncertainty of estimated by the deviating point of the regression fits and the linear fit of $\sqrt{M_{t}}$. The solid red line is the fitting curve using the two-gap model. The dashed and dash dotted lines show the contributions in the two-band model of the large gap and small gap, respectively. The dotted and solid blue lines present the one-gap and $d$-wave fit, respectively.}
\end{figure}


The $T$-dependence of the corrected values $H_{c1}$ applied along the $c$-axis is shown in Fig.\,3. The main features in Fig.\,3, $H_{c1}$-data, can be described in the following ways: (i) As the first step we compare our data to the $d$-wave and single-gap BCS theory under the weak-coupling approach, see dotted and solid blue lines. Indeed, both quantities lead to a rather different trend and show a systematic deviation from the data in the whole $T$-range below $T_{c}$. (ii) Then, the obtained experimental $H_{c1}$ data were analyzed by using the phenomenological $\alpha$-model, see dashed lines in Fig.\,3. The $T$-dependence of each energy gap for this model can be approximated as:~\cite{Carrington} $\Delta _{i}(T) = \Delta _{i}(0) {\tanh[1.82(1.018(\frac{T_{ci}}{T}-1))^{0.51}]}$, where $\Delta(0)$ is the maximum gap value at $T$ = 0. We adjust the temperature dependence by using the following expression:
\begin{equation}
 \frac{\lambda _{ab}^{-2}(T)}{\lambda _{ab}^{-2}(0)} = 1+\frac{1}{\pi}\int^{2 \pi}_0{ 2\int_{\Delta(T,\phi)}^{\infty}{\frac{\partial f}{\partial E} \frac{E dE d\phi}{\sqrt{E^2-\Delta^2(T,\phi)}}}},
\end{equation}
where $f$ is the Fermi function $ [ \exp( \beta E + 1 )]^{-1}$, $\varphi$ is the angle along the Fermi surface, $\beta$ = ($k_\textup{B}T)^{-1}$. The energy of the quasiparticles is given by $E$ = $[\epsilon^{2} + \Delta^{2}(t)]^{0.5}$, with $\epsilon$ being the energy of the normal electrons relative to the Fermi {level}, and where $\Delta(T,\phi)$ is the order parameter as function of temperature and angle. We used for the $s$-wave, $d$-wave the following expressions  $\Delta(T,\phi)=\Delta(T)$ and $\Delta(T,\phi)=\Delta(T) \cos(2 \theta)$. For the two-gap model, we calculated as:~\cite{Carrington}
 {\begin{equation}
\lambda _{ab}^{-2}(T) = r\lambda _{1}^{-2}(T) + (1-r)\lambda _{2}^{-2}(T),
\end{equation}}
where $0<r<1$. Equations (2) and (3) are used to introduce the two gaps and their appropriate weights. The best description of the experimental data is obtained using values of $\Delta_{1}$ = 2.2$\pm0.3$\,meV, $\Delta_{2}$ = 11.5 $\pm0.4$\,meV and $r$ = 0.45 and 0.35 for $x$ = 0.1 and 0.2 (see supplementary materials) respectively and represented by the solid red lines in Fig.\,3. The value of the gap amplitudes obtained for this material scales relatively well with its $T_{c}$ in light of the recent results for the Fe-based superconductors~\cite{Paglione2010}. From the present $H_{c1}$ results, (i) Two $s$-wave-like gaps are accounted for the $T$-dependence of the  $H_{c1}$; (ii) kinky structure in $H_{c1}(T)$; and (iii) the large gap value of F-doped is considerably larger than the BCS weak-coupling limit, these observations clearly show that there are no nodes in the SC energy gap indicating a multiband and nodeless superconductivity in the F-doped NdFeAsO.

\begin{figure}[b]
\includegraphics[width=21pc,clip]{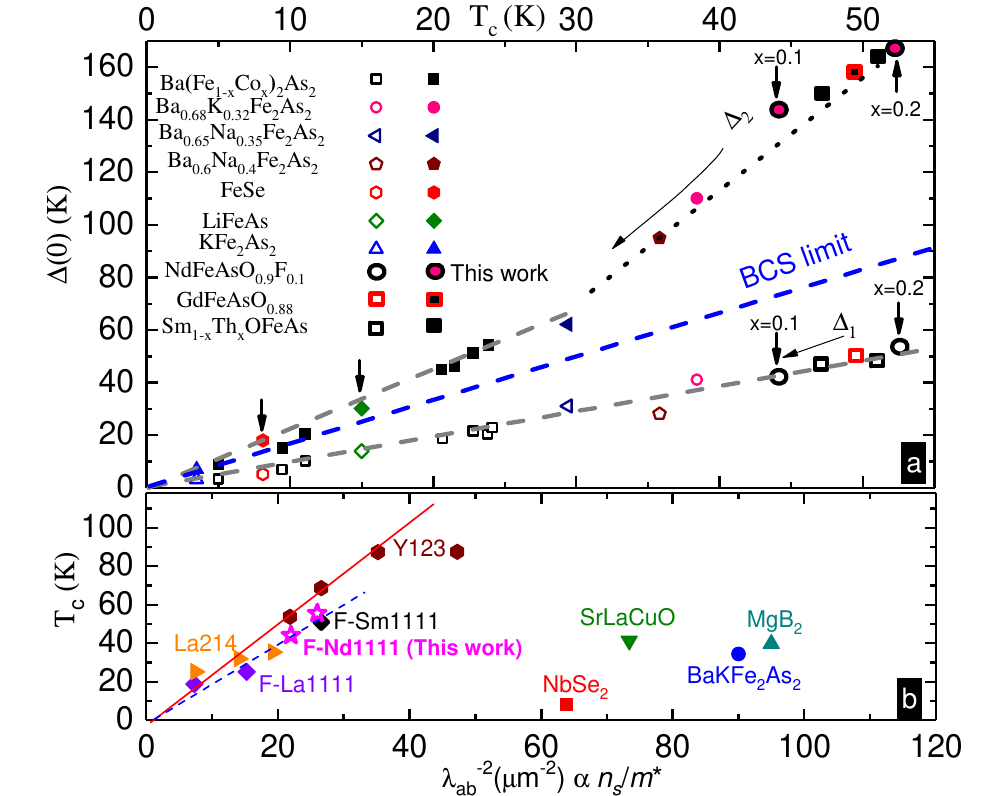}
\caption{(a)  The dependence of the large gap (solid symbols) and the small gap (open symbols) on the critical temperature NdFeAsO$_{0.9}$F$_{0.1}$ and NdFeAsO$_{0.8}$F$_{0.2}$  (this study) together with Ba$_{0.65}$Na$_{0.35}$Fe$_{2}$As$_{2}$,~\cite{Pramanik2011} Ba(Fe$_{1-x}$Co$_{x}$)$_{2}$As$_{2}$~\cite{hardy} for $0.05\leq x \geq0.146$, Ba$_{0.68}$K$_{0.32}$Fe$_{2}$As$_{2}$,~\cite{Popovich2010} Ba$_{0.6}$K$_{0.4}$Fe$_{2}$As$_{2}$,~\cite{CR} KFe$_{2}$As$_{2}$,~\cite{Hafiez2011} LiFeAs,~\cite{YS}, Ba$_{0.45}$K$_{0.55}$Fe$_{2}$As$_{2}$,~\cite{TSh} LaFeAsO$_{0.9}$F$_{0.1}$,~\cite{YaG}, Sm$_{1-x}$Th$_{x}$OFeAs~\cite{two} and FeSeTe~\cite{TK}. BCS limit is shown by the dash-blue line for comparison. Black lines are guidelines. The arrows show that both gap values are determined using the lower critical field. (b) The correlation between $T_{c}$ and the superfluid density $n_{s}/m^{*}$. The obtained $\lambda _{ab}(0)$ F-Nd1111 samples roughly follows the Uemura relation in agreement with F-LaFeAsO and F-SmFeAsO.}
\end{figure}

The large gap $\Delta_2$ has a higher value than the weak-coupling BCS (1.76$k_{B}T_{c}$) gap value, which reflects a tendency for strong coupling effects, while the smaller one $\Delta_1$ has a value lower than the BCS one. This is consistent with the theoretical constraints that one gap must be larger than the BCS gap and one smaller in a weakly coupled two-band superconductor~\cite{Saito}. This variation of the large gap from the BCS-limit could be caused by several reasons, such as (i) out-of-plane anisotropy of the order parameter discussed in\cite{Saito}, (ii) a complex and nontrivial in-plane angle distribution of the large gap in the $k$-space, (iii) a possible presence of a large gap splitting, (iv) a surface sensitivity of SC properties, (vi) a significant contribution of high-energy $\omega > \Delta$ pairs with $Re[\Delta(\omega)] > \Delta _{exp}$ (where $\Delta _{exp}$ is a gap edge of the Eliashberg function) accounted in bulk probes. As for the small gap, the extracted value lies well below the BCS limit and point to a nonzero interaction between the condensates. It is important to note thatthe simple $\alpha$-model that is not self-consistent, but is often used by experimentalists for fitting their thermodynamic data that deviate from the BCS predictions and for quantifying those deviations\cite{m8}. However, it is a matter of concern to be sure whether one, two or three bands can describe well our experimental data, since in the case of multiband superconductivity low-energy quasiparticle excitations can be always explained by the contribution from an electron group with a small gap. By complementing presented $H_{c1}$, the two-band model seems to be sufficient to describe the experimental temperature dependencies of SC parameters.


By summarizing the gap values determined by $H_{c1}$ data of the Nd-1111 sample, one may unreveal the influence of fluorine doping on the SC properties $[$Fig.\,4(a)$]$. The $\Delta_{2}$ values
are shown by solid symbols, and the $\Delta_{1}$ values are shown by open symbols. The data obtained earlier in the 122, 11, 1111 and 1111 systems are taken from~\cite{hardy,Popovich2010,Pramanik2011,Hafiez2011,TSh,YaG,two,m2}. As can be seen, the gap values differ for different compounds and  the small gap increases linearly with $T_{c}$. The family of 1111 superconductors with Gd, La, Ce, and Nd follow this tendency~\cite{two,Te}. It is clearly seen in Fig.\,4(a) that the larger gap, a tendency for strong-coupling effects, increases stronger than linear with $T_{c}$ for $T_{c}$ $\geq$ 30\,K. The larger gap converges to the BCS value, which has previously been reported also for Ba(Fe$_{1-x}$Co$_{x}$)$_{2}$As$_{2}$ and Mg$_{1-x}$Al$_{x}$B$_{2}$~\cite{hardy,Al}. By analogy, the difference in gap amplitude could be comes from the coupling strength and the interband scattering between the 2D $\sigma$ bands (which exhibit the larger gap) and the 3D $\pi$ bands are weak because of their difference in character. We believe that our data and analysis are fully consistent with superconductivity driven by interband coupling between the nested $\beta$ and $\gamma$ bands.

Another important result of our experiment is the absolute value of $\lambda _{ab}(0)$, which is independently deduced from the measured $H_{c1}(0)$. In principle, we estimated the penetration depth at low temperatures using the traditional Ginzburg-Landau (GL) theory, where $H_{c1}$ is given by: $\mu_{0}H_{c1}^{\parallel c} = (\phi$$_{0}/4\pi\lambda _{ab}^{2})\ln\kappa _{c}$, where $\phi$$_{0}$ is the magnetic-flux quantum $\phi$$_{0}$ = $h/e^{\ast}$ = 2.07 x 10$^{-7}$Oe cm$^{2}$, $\kappa _{c}$ =$\lambda _{ab}$/$\xi _{ab}$ is the GL parameter, which we obtained $\lambda _{ab}$(0) = 218(10), 209(10)\,nm for $x$ = 0.1, and 0.2, respectively. The value of $\lambda _{ab}^{-2}(0)$, or equivalently the condensed carrier density $n_{s}/m^{*}$ (superconducting carrier density/effective mass), allows us to check whether the well-known scaling behavior between $n_{s}/m^{*}$ and $T_{c}$ still works for the investigated system. In Fig.\,4(b), we summarize our results together with others 1111 system~\cite{A,B,C}, 122~\cite{122}, cuprates~\cite{C}, MgB$_{2}$~\cite{Mg}, and NbSe$_{2}$~\cite{Nb}. It is remarkable that the F-doped samples follow the Uemura plot consistent with the other F-La1111 and F-Sm1111, but is contrasted to the 122 and NbSe$_{2}$ systems which is quite far from the Uemura plot. In order to understand the discrepancy between the latter systems and the 1111 warrants further investigation.

In summary, we used a complementary experimental techniques to study high-quality fluorine-doped NdFeAsO single crystals and obtained consistent data on the structure of the SC order parameter. Our data extracted from the $T$-dependence of lower critical field are inconsistent with a single $s$-wave order parameter but clearly show the presence of two gaps without nodes.  A tendency to strong coupling for the larger gap is observed and the obtained gap values are consistent with those determined previously~\cite{Nd1,Nd2}. The penetration depth at $T$ = 0\,K confirms that the F-Nd1111 superconductors obey an Uemura-style relationship between $T_{c}$ and $n_{s}/m^{*}$. More interestingly, by comparing the $H_{c2}$ anisotropy of the crystals grown with various fluxes in previous studies, the anisotropy is typically around 5$\pm$1. Based on our results and on the results of other studies, this small anisotropy seems to be universal features of the Fe-based superconductors and can be considered as arguments in support of the common multiband scenario proposed for FeAs-based superconductors~\cite{two1,two2,two3,P,sm}.

We are grateful for discussions with D. V. Efremov, S.-L. Drechsler, and V. V. Pudalov and support from DFG, Deutsche Forschungsgemeinschaft through MO 3014/1-1.


\begin{thebibliography}{100}

\bibitem{ZA}  Z.-A. Ren, W. Lu, J. Yang, W. Yi, X.-L. Shen, Z.-C. Li, G.-C. Che, X.-L. Dong, L.-L. Sun, F. Zhou, Z.-X. Zhao, Chin. Phys. Lett. \textbf{25}, 2215 (2008).

\bibitem{ZA1} Y. Kamihara, T. Watanabe, M. Hirano, and H. Hosono, J. Am. Chem. Soc. \textbf{130}, 3296 (2008).

\bibitem{ZA3} G. R. Stewart, Rev. Mod. Phys. 83, 1589 (2011).

\bibitem{ZA4} X. Chen, P. Dai, D. Feng, T. Xiang, and F. C. Zhang, Nat. Sci. Rev. \textbf{1}, 371 (2014).

\bibitem{Johnston2010} D. C. Johnston, Advances in Physics \textbf{59}, 803 (2010).

\bibitem{Paglione2010} J. Paglione and R. L. Greene, Nature Physics \textbf{6}, 645, (2010).

\bibitem{gap} G. F. Chen {\it et al.}, Phys. Rev. Lett. 100, 247002 (2008).

\bibitem{gap1} X. H. Chen {\it et al.}, Nature (London) 453, 761 (2008).

\bibitem{SC} Wang, F. \& Lee, D.-H.  Science \textbf{332}, 200-2004 (2011).

\bibitem{Hir} P. J. Hirschfeld, M. M. Korshunov, \& I. I. Mazin,  Rep. Prog. Phys. \textbf{74}, 124508 (2011).

\bibitem{Hir2} A. Chubukov \& P. J. Hirschfeld, Physics Today, \textbf{68}, 46 (2015).

\bibitem{Cup} C. C. Tsuei \& J. R. Kirtley, Rev. Mod. Phys. \textbf{72}, 969-1016 (2000).

\bibitem{two} T. E. Kuzmicheva, S. A. Kuzmichev, K. S. Pervakov, V. M. Pudalov, and N. D. Zhigadlo, Phys. Rev. B \textbf{95}, 094507 (2017).

\bibitem{two1} L. Malone, J. D. Fletcher, A. Serafin, and A. Carrington, N. D. Zhigadlo, Z. Bukowski, S. Katrych, and J. Karpinski, Phys. Rev. B \textbf{79}, 140501(R), (2009).

\bibitem{two2} D. Daghero, M. Tortello, R. S. Gonnelli, V. A. Stepanov, N. D. Zhigadlo, and J. Karpinski, Phys. Rev. B \textbf{80}, 060502(R), (2009).

\bibitem{two3} Y.-L. Wang, L. Shan, L. Fang, P. Cheng, C. Ren and H.-H. Wen, Supercond. Sci. Technol. \textbf{22}, 015018 (2009).

\bibitem{Nd1} C. Martin, M. E. Tillman, H. Kim, M. A. Tanatar, S. K. Kim, A. Kreyssig, R. T. Gordon, M. D. Vannette, S. Nandi, V. G. Kogan, S. L. Budko, P. C. Canfield, A. I. Goldman, and R. Prozorov, Phys. Rev. Lett. \textbf{102}, 247002 (2009).

\bibitem{Nd2} P. Samuely, P. Szabo, Z. Pribulova, M E Tillman, S. L. Budko and P. C. Canfield, Supercond. Sci. Technol. \textbf{22}, 014003 (2009).

\bibitem{P} I. I. Mazin, D. J. Singh, M. D. Johannes, and M. H. Du, Phys. Rev. Lett. \textbf{101}, 057003 (2008).

\bibitem{P1} K. Kuroki, S. Onari, R. Arita, H. Usui, Y. Tanaka, H. Kontani, and H. Aoki, Phys. Rev. Lett. \textbf{101}, 087004 (2008).

\bibitem{P2} Xi Dai, Zhong Fang, Yi Zhou, and F. C. Zhang, Phys. Rev. Lett. \textbf{101}, 057008 (2008).

\bibitem{P3} P. A. Lee and X. G. Wen, Phys. Rev. B \textbf{78}, 144517 (2008).

\bibitem{P4} Xi Dai, Zhong Fang, Yi Zhou, and Fu-Chun Zhang, Phys. Rev. Lett. \textbf{101}, 057008 (2008).



\bibitem{m2} M. Abdel-Hafiez, J. Ge, A. N. Vasiliev, D. A. Chareev, J. Van de Vondel, V. V. Moshchalkov, and A. V. Silhanek, Phys. Rev. B \textbf{88}, 174512 (2013).

\bibitem{m3} M. Abdel-Hafiez, P. J. Pereira, S. A. Kuzmichev, T. E. Kuzmicheva, V. M. Pudalov, L. Harnagea, A. A. Kordyuk, A. V. Silhanek, V. V. Moshchalkov, B. Shen, Hai-Hu Wen, A. N. Vasiliev, and X.-J. Chen, Phys. Rev. B \textbf{90}, 054524 (2014).

\bibitem{m4} M.Abdel-Hafiez, Y. Zhang, Z. He, J. Zhao, C. Bergmann, C. Krellner, C. Duan, X. Lu, H. Luo, P. Dai, and X..J. Chen, Phys. Rev. B \textbf{91}, 024510 (2015).

\bibitem{CR} C. Ren, Z.-S. Wang, H.-Q. Luo, H. Yang, L. Shan, H.-H. Wen, Phys. Rev. Lett. \textbf{101 } 257006, (2008).

\bibitem{REN} C. Ren, Z-S. Wang, H. Yang, X. Zhu, L. Fang, G. Mu, L. Shan, and H.-H Wen, ArXiv:0804.1726.

\bibitem{Martin} C. Martin, R. T. Gordon, M. A. Tanatar, H. Kim, N. Ni, S. L. Budko, P. C. Canfield, H. Luo, H. H. Wen, Z. Wang, A. B. Vorontsov, V. G. Kogan, and R. Prozorov  Phys. Rev. B \textbf{80}, 020501(R) (2009).

\bibitem{RG} R. T. Gordon, {\it et al.}, Phys. Rev. Lett. \textbf{102}, 127004 (2009).

\bibitem{Nd3} T. Kondo, A. F. Santander-Syro, O. Copie, Chang Liu, M. E. Tillman, E. D. Mun, J. Schmalian, S. L. Bud'ko, M. A. Tanatar, P. C. Canfield, and A. Kaminski, Phys. Rev. Lett. \textbf{101}, 147003 (2008).

\bibitem{Nd4} X L Wang, S X Dou, Z. Ren, W. Yi, Z. Li, Z. Zhao and S. Lee, J. Phys. Condens. Matter \textbf{21}, 205701 (2009).

\bibitem{1a} Y. Chen, J. W. Lynn, J. Li, G. Li, G. F. Chen, J. L. Luo, N. L. Wang, P. Dai, C. dela Cruz, and H. A. Mook, Phys. Rev. B \textbf{78}, 064515 (2008).


\bibitem{1b} W. Tian, {\it et al.}, Phys. Rev. B \textbf{82}, 060514(R) (2010).

\bibitem{NR} N. R. Werthamer, E. Helfand, and P. C. Hohenberg, Phys. Rev. \textbf{147}, 295 (1966).

\bibitem{Jia2008} Y. Jia, P. Cheng, L. Fang, H. Luo, H. Yang, C. ren, L. Shan, C. Gu, H.-H. Wen, Appl. Phys. Lett. \textbf{93}, 032503 (2008).

\bibitem{sm} N. D. Zhigadlo, S. Weyeneth, S. Katrych, P. J. W. Moll, K. Rogacki, S. Bosma, R. Puzniak, J. Karpinski, and B. Batlogg, Phys. Rev. B. \textbf{86}, 214509 (2012).

\bibitem{Moshchalkov} V. V. Moshchalkov, J. Y. Henry, C. Marin, J. Rossat-Mignod, J. F. Jacquot, Physica C \textbf{175}, 407 (1991).

\bibitem{Angst} M. Angst, R. Puzniak, A. Wisniewski, J. Jun, S. M. Kazakov, J. Karpinski, J. Roos, and H. Keller, Phys. Rev. Lett. \textbf{88}, 167004 (2002).

\bibitem{Brandt} E. H. Brandt, Phys. Rev. B \textbf{60}, 11939 (1999).

\bibitem{Carrington} A. Carrington and F. Manzano, Physica C \textbf{385}, 205 (2003).

\bibitem{Saito} T. Saito, S. Onari, and H. Kontani, Phys. Rev. B \textbf{88}, 045115 (2013).




\bibitem{m8} D. C. Johnston, Supercond. Sci. Technol. \textbf{26}, 115011 (2013).

\bibitem{Al} R. Gonnelli, A. Calzolari A. and D. Daghero, J. Supercond. Nov. Magn., \textbf{20}, 555 (2007).

\bibitem{hardy} F. Hardy, {\it et al.}, Europhys. Lett. \textbf{91}, 47008 (2010).

\bibitem{Popovich2010} P. Popovich, A. V. Boris, O. V. Dolgov, A. A. Golubov, D. L. Sun, C. T. Lin, R. K. Kremer, and B. Keimer, Phys. Rev. Lett. \textbf{105}, 027003 (2010).

\bibitem{Pramanik2011} A. K. Pramanik, M. Abdel-Hafiez, S. Aswartham, A. U. B. Wolter, S. Wurmehl, V. Kataev, and B. B\"{u}chner, Phys. Rev. B \textbf{84}, 064525 (2011).

\bibitem{Hafiez2011} M. Abdel-Hafiez, {\it et al.}, Phys. Rev. B \textbf{85}, 134533 (2012).

\bibitem{YS} Y. Song, J. Ghim, J. Yoon, K. Lee, M. Jung, H. Ji, J. Shim, Y. Kwon, Europhys. Lett. \textbf{94}, 57008 (2011).

\bibitem{TSh} T. Shibauchi, K. Hashimoto, R. Okazaki, Y. Matsuda Physica C \textbf{469}, 590 (2009).

\bibitem{YaG} Ya. G. Ponomarev, S. A. Kuzmichev, M. G. Mikheev, M. V. Sudakova, S. N. Tchesnokov, O. S. Volkova, A. N. Vasiliev, T. H\"{a}nke, C. Hess, G. Behr, R. Klingeler, and B. B\"{u}chner, Phys. Rev. B \textbf{97}, 224517 (2009).

\bibitem{TK}T. Kato, Y. Mizuguchi, H. Nakamura, T. Machida, H. Sakata, and Y. Takano, Phys. Rev. B \textbf{80}, 180507(R) (2009).

\bibitem{Te} T. E. Kuzmicheva, S. A. Kuzmichev, M. G. Mikheev, Ya. G. Ponomarev, S. N. Tchesnokov, V. M. Pudalov, E. P. Khlybov, and N. D. Zhigadlo, Physics-Uspekhi \textbf{57}, 819 (2014).

\bibitem{A} A. J. Drew, F. L. Pratt, T. Lancaster, S. J. Blundell, P. J. Baker, R. H. Liu, G. Wu, X. H. Chen, I. Watanabe, V. K. Malik, A. Dubroka, K. W. Kim, M. R\"{o}ssle, and C. Bernhard, Phys. Rev. Lett. \textbf{101}, 097010 (2008).

\bibitem{B} G. Mu, X. Zhu, L. Fang, L. Shan, C. Ren, H.-H. Wen Chin. Phys. Lett. \textbf{25}, 2221 (2008).

\bibitem{122} C. Ren, Z. Wang, H. Luo, H. Yang, L. Shan, and H.-H. Wen, Phys. Rev. Lett. \textbf{101}, 257006 (2008).

\bibitem{C} H. Luetkens, H.-H. Klauss, R. Khasanov, A. Amato, R. Klingeler, I. Hellmann, N. Leps, A. Kondrat, C. Hess, A. Köhler, G. Behr, J. Werner, and B. B\"{u}chner, Phys. Rev. Lett. \textbf{101}, 097009 (2008).

\bibitem{Mg} F. Manzano, A. Carrington, N. E. Hussey, S. Lee, A. Yamamoto, and S. Tajima, Phys. Rev. Lett. \textbf{88}, 047002 (2002).

\bibitem{Nb} J. D. Fletcher, A. Carrington, P. Diener, P. Rodière, J. P. Brison, R. Prozorov, T. Olheiser, and R. W. Giannetta, Phys. Rev. Lett. \textbf{98}, 057003 (2007).


\end{thebibliography}
\end{document}